\documentclass[a4paper,11pt]{article}
\usepackage{jcappub}
\usepackage[T1]{fontenc}
\usepackage{caption}
\usepackage{subcaption}
\usepackage{hyperref}
\usepackage{amsmath, amssymb, graphics}
\usepackage{mathtools}
\usepackage{tablefootnote}
\usepackage{soul}
\usepackage{ulem}
\usepackage{color}
\usepackage{float}
\usepackage{makecell}
\usepackage{booktabs}
\usepackage[usenames,dvipsnames,svgnames,table]{xcolor}
\usepackage{indentfirst}
\usepackage{leftidx}
\usepackage{multirow}

\definecolor{LightCyan}{rgb}{0.88,1,1}
\newcolumntype{a}{>{\columncolor{LightCyan}}c}

\newcommand{\apjl}{Astrophys. J. Lett.}
\newcommand{\apj}{Astrophys. J.}

\newcommand{\aap}{Astron. Astrophys.}

\newcommand{\mnras}{Mon. Not. Roy. Astron. Soc.}

\newcommand{\jcap}{JCAP}

\usepackage{xspace}

\newcommand{\hmsun}{$h^{-1}M_\odot$\xspace}

\title{Cosmic background neutrinos deflected by gravity: DEMNUni simulation analysis}

\author[1]{Beatriz Hern\'andez-Molinero,}
\author[2]{Carmelita Carbone,}
\author[3,4]{Raul Jimenez,}
\author[1,4]{Carlos Pe\~na Garay,}

\affiliation[1]{Laboratorio Subterr\'aneo de Canfranc, 22880 - Estaci\'on de Canfranc, Huesca, Spain.}
\affiliation[2]{INAF – Istituto di Astrofisica Spaziale e Fisica cosmica di Milano (IASF-MI), Via Alfonso Corti 12, I-20133 Milano, Italy}
\affiliation[3]{ICCUB, University of Barcelona, Marti  i Franques 1, E-08028 Barcelona, Spain.}
\affiliation[4]{Instituci\`o Catalana de Recerca i Estudis Avan\c{c}ats, Pg. Lluis Companys 23, Barcelona, E-08010, Spain.}
\affiliation[4]{I2SysBio, CSIC-University of Valencia, 46071 - Valencia, Spain.}

\emailAdd{bhernandez@lsc-canfranc.es}
\emailAdd{carmelita.carbone@inaf.it}
\emailAdd{raul.jimenez@icc.ub.edu}
\emailAdd{cpenya@lsc-canfranc.es}

\abstract{
The local supercluster acts as a gravity deflection source for cosmic background neutrinos. This deflection by gravity changes the neutrino helicity and therefore has important consequences for ground based tritium capture experiments aimed at determining if the neutrino is Dirac or Majorana. Here we explore the deflection effect of the local supercluster using the higher resolution  DEMNUni simulation suite and reaffirm our previous results. We show that the lightest neutrinos are ultra-relativistic enough to suffer little deflection by gravity and at the same time not relativistic enough to achieve the
same capture rate for Dirac and Majorana cases. This means that the capture rate in
Ptolemy-like experiments will be sensitive to the neutrino nature and that gravity deflection enlarges the difference between Majorana and Dirac rates. Moreover, using the relation between mass and momentum of the neutrinos frozen Fermi-Dirac distribution, we are able to calculate the deflection angle for different neutrino masses from the same set of neutrinos obtained from the simulation. Doing so, we provide a formula to compute the deflection angle for any neutrino mass, such that when cosmology detects an absolute neutrino mass, precise predictions can be made for tritium ground-based detectors on Earth aimed to determine neutrinos nature.}

\begin{document}
\maketitle	

\section{Introduction}
One second after the Big Bang, neutrinos decoupled from matter and streamed freely from gravity potential wells with a frozen Fermi-Dirac distribution. After this event, the neutrino content decoheres in mass eigenstates because neutrino energy and density redshift. The estimated temperature of the Fermi-Dirac distribution is $1.95 K$~\cite{lesgourges, RouletBook} nowadays. Because this distribution is frozen, all neutrinos, independently of their masses, have the same momentum distribution at present. However, the distribution of velocities is not the same because momentum is related to velocity by neutrino mass. 

Neutrinos, as no-charged fermions, can be either Dirac or Majorana. The current experimental route to discover their nature is neutrinoless double-beta decay~\cite{Goeppert-Mayer,furry}. However, cosmological observations already tell us that the sum of neutrino masses is small ($< 0.1$eV)~\cite{Planck:2018vyg,shapefit} and their hierarchy is normal~\cite{hierarchy,hierarchy2}; this would imply that double-beta decay experiments have to be in the several tons range, which is challenging. It is therefore crucial to look at cosmological signals. Experiments sensitive to the cosmic neutrino background ($C_\nu B$) will detect only left-handed neutrinos if neutrinos are Dirac particles but they will be sensitive to both left- and right-handed neutrinos if those are a Majorana particle. 

As the Universe has been cooling, neutrinos have become less relativistic and helicity and chirality states do not coincide. So, if neutrinos deflect due to the gravitational field produced by clustered structures, then the helicity content can be flipped. This has no consequences for the Majorana case as $C_\nu B$ detectors are sensitive to both active states but, for Dirac cases, gravity could redistribute the abundances decreasing left-handed neutrinos and increasing right-handed neutrinos. Gravity could reduce capture rates for Dirac neutrinos enhancing the difference between Dirac and Majorana cases.

In our previous work~\cite{Hernandez-Molinero:2022zoo}, we demonstrated that the neutrino velocity field is locally deflected over time and, indeed, the helicity content of Dirac neutrinos is modified. In that work, we assumed that different neutrinos mass eigenstates were equally populated, but, in reality, this is not the case: different neutrinos masses have different velocity fields so will suffer different deflection when falling into the same gravitational potential.
In principle, one could produce large-scale structure simulations for many different neutrino masses. However, high-resolution numerical simulations of large-scale structure that include cold dark matter and massive neutrinos have big computational requirements so it is not viable to make simulations for specific neutrino masses. Specially challenging is to simulate very low, $< 0.01$ eV, massive neutrinos as they are nearly relativistic, but yet are the ones that contribute the most to scattering in ground-based tritium-capture experiments. 

Here, we present a solution to this problem: using one simulation with massive neutrinos of a specific mass we come up with a formula that provides the mean cosine of the deflection angle observable on Earth as a function of the neutrino mass. Knowing the deflection angle, left- and right-handed neutrinos density is a straightforward calculation. To reproduce the deflection that a neutrino travelling inside our Local Supercluster may suffer, we have calculated deflection angles inside dark matter haloes available in the simulation with masses bigger than $10^{14}\, M_\odot$.

The Milky Way is in the plane of the local super-cluster~\cite{supercluster}, which is a collection of galaxies extending, and including, the local group and the Virgo cluster to up to about 18 Mpc. This local supercluster also contains a local void that is fairly empty of large massive galaxies like the Milky Way and Andromeda. It is this big structure that contributes most to the deflection of the cosmic neutrino background ($C_{\nu}B$) on Earth.

We update the calculations in~\cite{Hernandez-Molinero:2022zoo} computing the helicity states density for a general mass eigenstate. Coming up with a general formula is needed because the actual mass of neutrinos remains unknown. When the neutrino mass eigenstates will be determined, calculating expected rates in experiments based on neutrinos capture could be done directly using our formula.

This paper is organised as follows. In \S~\ref{sec:deflection} we compute the deflection angle for the different velocities. In \S~\ref{sec:capture} we provide the general formula to compute the capture rates in tritium experiments for any neutrino mass. We summarise our results in \S~\ref{sec:conclusions}.

\section{Deflection angle distribution reconstruction}
\label{sec:deflection}

As the Universe is clustered and non-linear gravitational structures exist, neutrinos will suffer a deflection angle when falling into a gravitational potential. This means that the helicity content of the C$_\nu$B is modified by gravity in the case that neutrinos are Dirac particles, which results in a decrease of the Dirac capture rates for experiments based on cosmic background neutrinos capture as Ptolemy~\cite{PTolemy,Hernandez-Molinero:2022zoo}. Capture rates depend on neutrino mass and neutrinos helicity states densities. But it is also important to consider that neutrinos densities are not independent of neutrinos masses. These experiments will be sensitive to deflection in the solar environment so it is needed to compute the deflection constraining the simulation trying to reproduce the Milky Way gravitational potential (or cosmological conditions).

Overcoming both requirements is a difficult task because performing simulations ad-hoc requires a high computational effort as following ultra-relativistic (low mass) neutrinos is a challenge due to their fast velocities. We present here a method to calculate deflection angles for different neutrino mass eigenstates without the necessity of new simulations for lighter neutrinos (<0.1 eV). In this analysis, we have used the ``Dark Energy and Massive Neutrino Universe'' (\href{https://www.researchgate.net/project/DEMN-Universe-DEMNUni}{DEMNUni}) suite of large N-body simulations~\cite{carbone_2016_demnuni}. The DEMNUni simulations have been produced with the aim of investigating the large-scale structure of the Universe in the presence of massive neutrinos and dynamical DE, and they were conceived for the nonlinear analysis and modelling of different probes, including dark matter, halo, and galaxy clustering~\cite{castorina_2015,moresco_2016,zennaro_2018,ruggeri_2018,bel_2019,parimbelli_2021,parimbelli_2022,Guidi_2022, SHAM-Carella_in_prep}, CMB lensing, Sunyaev-Zel'dovich and integrated Sachs-Wolfe effects ~\cite{carbone_2016_demnuni,roncarelli_2015,fabbian_2018}, cosmic void statistics~\cite{kreisch_2019,schuster_2019,verza_2019,verza_2022}, as well as cross-correlations among these probes~\cite{Vielzeuf2022_inprep,Cuozzo2022_inprep}.

The DEMNUni simulations combine a good mass resolution with a large volume to include perturbations both at large and small scales. They are characterised by a comoving volume of $(2 \: h^{-1}\mathrm{Gpc})^3$ filled with $2048^3$ dark matter particles and, when present, $2048^3$ neutrino particles.
The DEMNUni set is composed by 15 simulations, implementing the cosmological constant and 4 dynamical DE-EoS for each of the total neutrino masses considered in the degenerate mass scenario with three active neutrinos, i.e. $\sum m_\nu = 0,\, 0.16,\, 0.32\, {\rm eV}$. In this analysis we used the case with $\sum m_\nu = 0.16\, {\rm eV.}$

The other cosmological parameters of the simulations are based on a Planck 2013~\cite{planck2013} LCDM reference cosmology (with massless neutrinos), in particular: $n_{\rm s}=0.96$, $A_{\rm s}=2.1265 \times 10^{-9}$, $h=H_0/[100\, 
{\rm km} \, s^{-1}{\rm Mpc}^{-1}]=0.67$, $\Omega_{\rm b}=0.05$, and $\Omega_{\rm m}=\Omega_{\rm CDM} + \Omega_{\rm b} + \Omega_\nu =0.32$; $H_0$ is the Hubble constant at the present time, $n_{\rm s}$ is the spectral index of the initial scalar perturbations, $A_{\rm s}$ is the scalar amplitude, $\Omega_{\rm b}$ the baryon density parameter, $\Omega_{\rm m}$ is the total matter density parameter, $\Omega_{\rm CDM}$ the cold dark matter density parameter, and $\Omega_\nu$ the neutrino density parameter. In the presence of massive neutrinos, $\Omega_{\rm b}$ and $\Omega_{\rm m}$ are kept fixed to the above values, while $\Omega_{\rm CDM}$ is changed accordingly. Tab.~\ref{tab:neutrino_params} summarises the masses of the CDM and neutrino particles together with the neutrino fraction $f_\nu \equiv \Omega_\nu / \Omega_{\rm m}$. 
\begin{table}[t]
\centering
\vspace{2ex}
\setlength{\tabcolsep}{0.7em}
\begin{tabular}{cccc}
\toprule
$\sum m_\nu$  [eV] & $f_\nu$ & $m_{\rm p}^{\rm CDM}$  [\hmsun] & $m_{\rm p}^\nu$  [\hmsun] \\
\midrule
\hline 
0     & 0 & $8.27\times 10^{10}$ & 0 \\
0.16  & 0.012 & $8.17\times 10^{10}$ & $0.99\times 10^9$ \\
0.32 & 0.024 & $8.07\times 10^{10}$ & $1.98\times 10^9$ \\
\noalign{\vspace{1ex}}
\hline 
\bottomrule
\end{tabular}
\caption{Summary of particle masses and neutrino fractions implemented in the DEMNUni simulations. The first column shows the total neutrino mass, the second the fraction of neutrinos and matter density parameters, and the last two columns show the corresponding mass of CDM and neutrino particles implemented in the simulations. 
}
\label{tab:neutrino_params}
\end{table}
Dark matter haloes are identified using a friends-of-friends (FoF) algorithm~\cite{davis_1985_fof} applied to dark matter particles, with a minimum number of particles fixed to 32, corresponding to a mass of $\sim 2.5 \times 10^{12} h^{-1}M_{\odot}$, and a linking length of 0.2 times the mean particle separation. FoF haloes are further processed with the {\sc subfind} algorithm~\cite{springel_2001_gadeget,dolang_2009_gadget} to produce subhalo catalogues. With this procedure, some of the initial FoF parent haloes are split into multiple substructures (see~\citep[][]{parimbelli_2022} for more details on the DEMNUni simulations). 
In the following with the term ``halo'' we will refer to the objects identified by the {\sc subfind} algorithm.

The deflection angle is calculated between neutrinos velocities at $z=3$, which is enough, and at $z=0$ as done in~\cite{Hernandez-Molinero:2022zoo}. 

\begin{figure}
    \centering
    \includegraphics[width=0.8\textwidth]{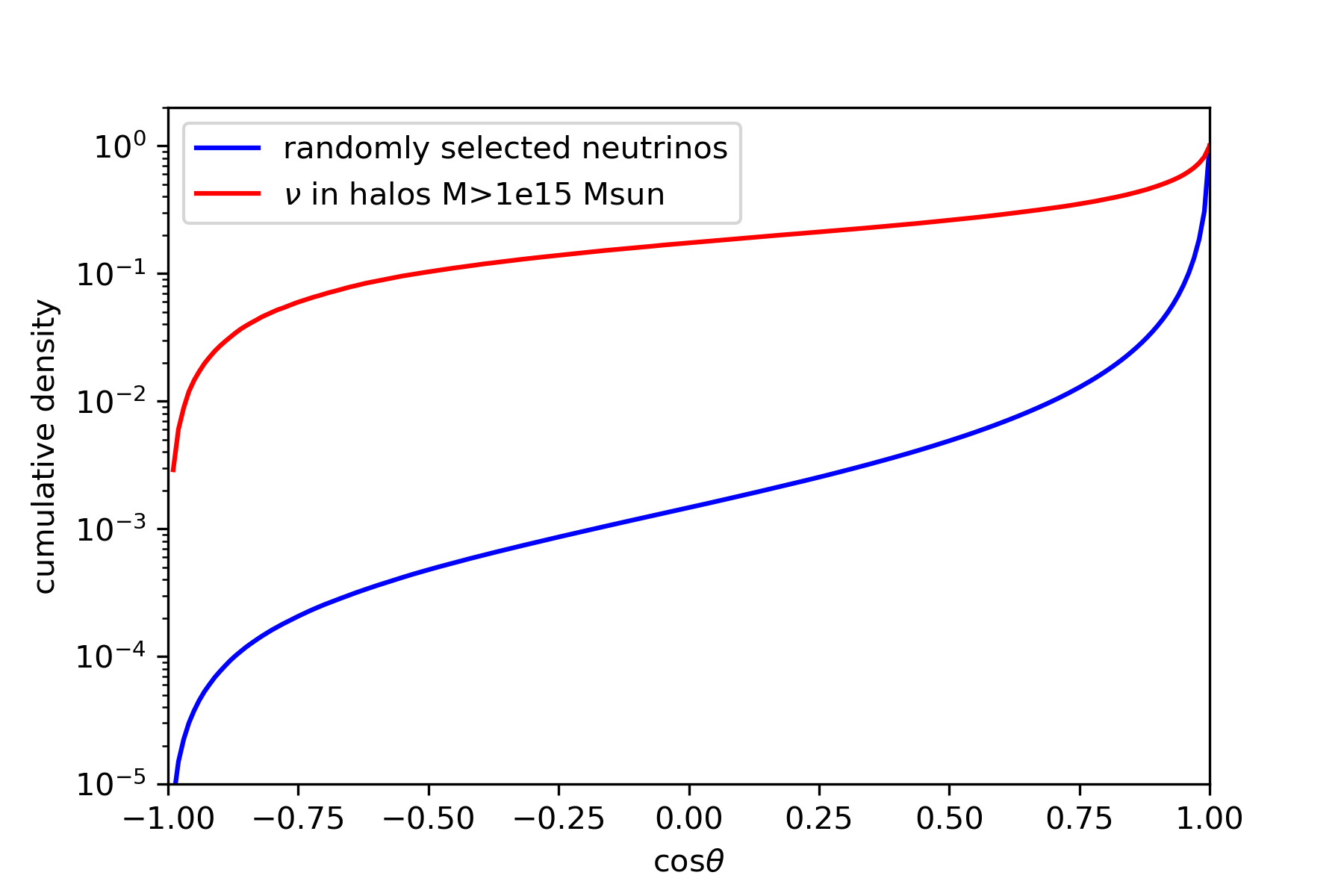}
    \caption{Deflection angle comparison between all neutrinos found in the simulation in blue versus neutrinos found inside haloes that satisfy $M>10^{15}\,M_\odot$ within 1.5 Mpc in red.}
    \label{fig:angle deflection distribution}
\end{figure}

In this work, we compute the dependence of the deflection angle distribution on the velocity distribution which is a function of the neutrino mass. For the simulation which we are working on, the obtained deflection angle distribution depends on the neutrino velocities distribution of this simulation, which is fixed at $z=0$ as a Fermi-Dirac distribution characterised by a temperature $T_\nu=1.95 \,K$ and a mass of 53 meV, i.e., neutrinos velocities follow the following distribution
\begin{equation}
    \frac{df^{FD}}{d\beta}=\frac{df^{FD}}{dq}\frac{dq}{d\beta} = \frac{4\pi}{(2\pi)^3}\frac{q(\beta)^2}{1+\exp(q(\beta)/T_\nu)}\frac{m}{\sqrt{(1-\beta^2)^3}},
\end{equation}
where $q$ stands for the neutrino momentum.

Secondly, to reproduce the solar environment, we focus on neutrinos inside dark matter haloes with masses larger than $10^{15}\, M_\odot$; with this constraint we get a mean halo mass of $1.4\,10^{15}\, M_\odot$. Neutrinos that may be detected by experiments on Earth have suffered deflections inside our Local Supercluster. Virgo Supercluster mass is about $1.48\,10^{15}\, M_\odot$, this is the reason why we perform the simulation analysis constraining to the most massive haloes. As a result of this constraint, we work with around 120000 neutrinos found in 1600 haloes. In Fig.~\ref{fig:angle deflection distribution} we compare the deflection of neutrinos in the most massive haloes with the deflection of all neutrinos of the simulation. It is clear, once again, that gravity modifies the neutrinos velocity field.

\begin{figure}
    \centering
    \begin{subfigure}[b]{0.8\textwidth}
    \centering
    \includegraphics[width=\textwidth]{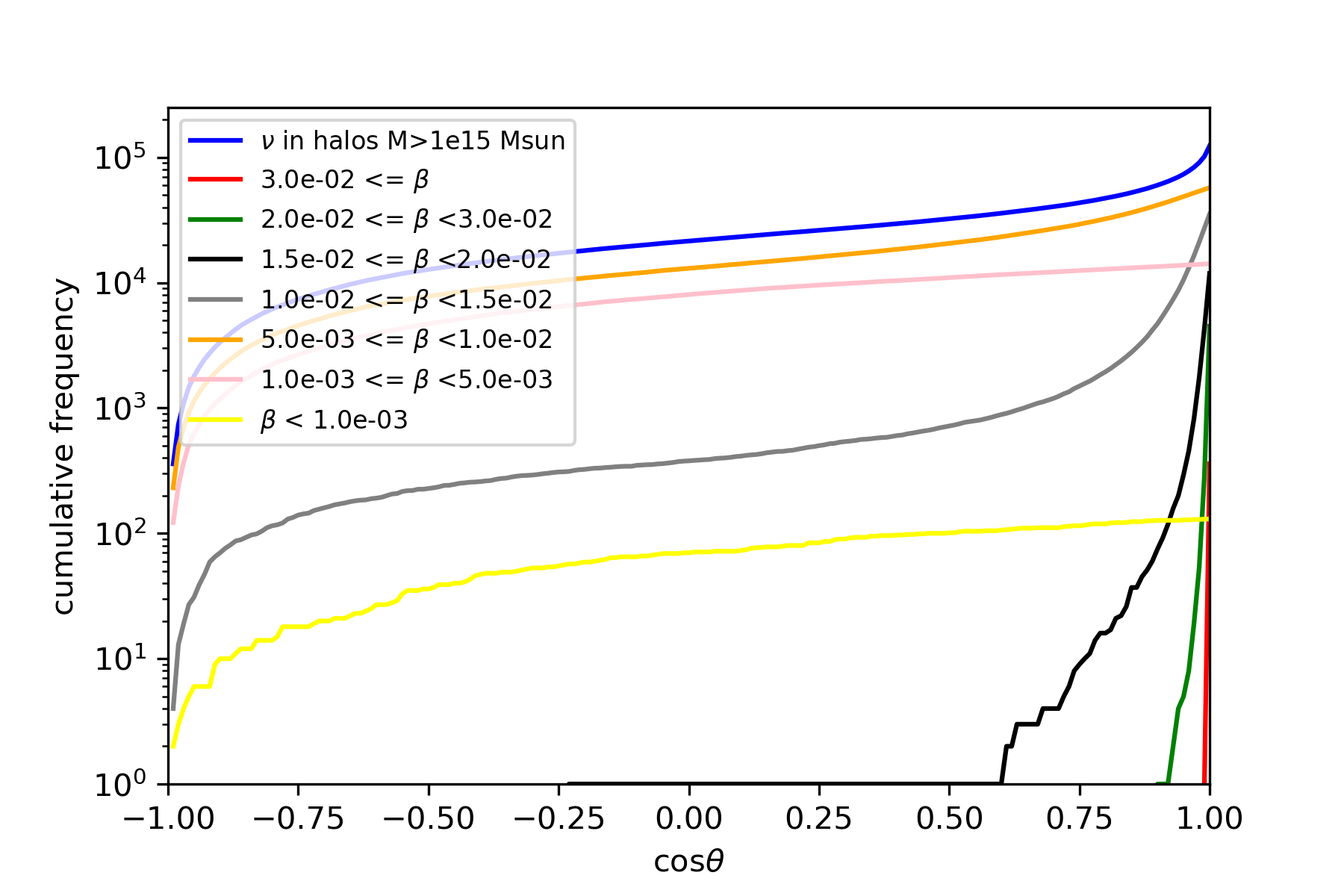}
    \caption{frequency}
    \end{subfigure}
    \hfill
    \begin{subfigure}[b]{0.8\textwidth}
    \centering
    \includegraphics[width=\textwidth]{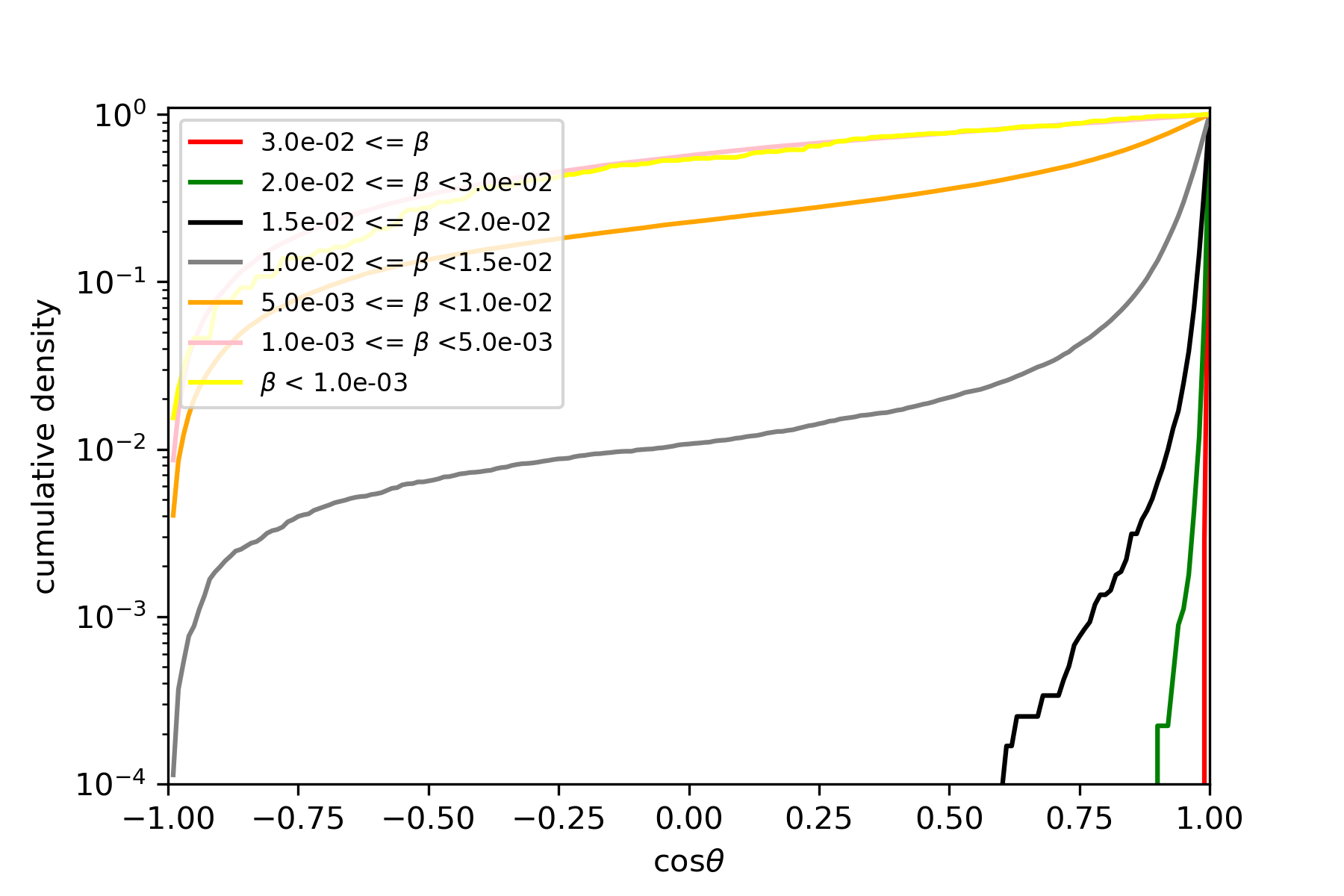}
    \caption{density}
    \label{fig:angle deflection beta ranges density}
    \end{subfigure}
    \caption{Deflection angle distribution in different beta ranges calculated over neutrinos found inside haloes that satisfy $M>10^{15}\,M_\odot$.}
    \label{fig:angle deflection beta ranges}
\end{figure}

Deflection angle distributions, as the one presented in Fig.~\ref{fig:angle deflection distribution}, depend on the velocity distribution as we were saying. In order to get deflection angle distributions for other neutrino masses without the need to run new simulations, we have used the one calculated from DEMNUni simulations with $\sum m_\nu = 0.16 \,{\rm eV}$ to reconstruct them. To do it, we have divided the angle distribution into different neutrino velocity ranges and have normalised the distributions for neutrinos with velocity in that range (Fig.~\ref{fig:angle deflection beta ranges}). As a result, we get a collection of normalised distributions that represent the deflection angle that neutrinos, of specific velocities, will suffer in our gravitational local environment.

Normalised distributions for different beta ranges (Fig.~\ref{fig:angle deflection beta ranges density}) will contribute differently to the total deflection angle distribution as a function of the neutrino mass. To get the total deflection angle distribution for a specific neutrino mass, we sum up normalised partial angle distributions weighted by the number of neutrinos in each beta range obtained from the Fermi-Dirac distribution for that specific neutrino mass (Eq.~\ref{eq: weigths})
\begin{equation}
    P(\beta_0<\beta<\beta_1)=\int_{\beta_0}^{\beta_1}{\frac{df^{FD}}{d\beta}d\beta}= \frac{4\pi}{(2\pi)^3}\int_{\beta_0}^{\beta_1}{\frac{q(\beta)^2}{1+\exp(q(\beta)/T_\nu)}\frac{m}{\sqrt{(1-\beta^2)^3}}d\beta}
    \label{eq: weigths}
\end{equation}

 We continue to use our reference case from previous work~\cite{Hernandez-Molinero:2022zoo}, namely, 
\begin{equation}
\begin{aligned}
    m_1 &= 1 \ \text{meV} \\
    m_2 &= 8 \ \text{meV} \\
    m_3 &= 50 \ \text{meV} \\
\end{aligned}
\label{eq:neutrino masses}
\end{equation}
that fulfil normal ordering and the following conditions
\begin{subequations}
    \begin{equation}
        \delta m ^2 = m_2^2 - m_1^2 = 7.5\times10^{-5}\,\text{eV}^2
        \label{mass_condition1}
    \end{equation}
    \begin{equation}
        \Delta m^2 = |m_3^2 - (m_2^2 + m_1^2)/2| = 2.5\times10^{-3}\,\text{eV}^2
    \label{mass_condition2}
    \end{equation}
\end{subequations}.

\begin{figure}
\centering
\includegraphics[width=0.8\textwidth]{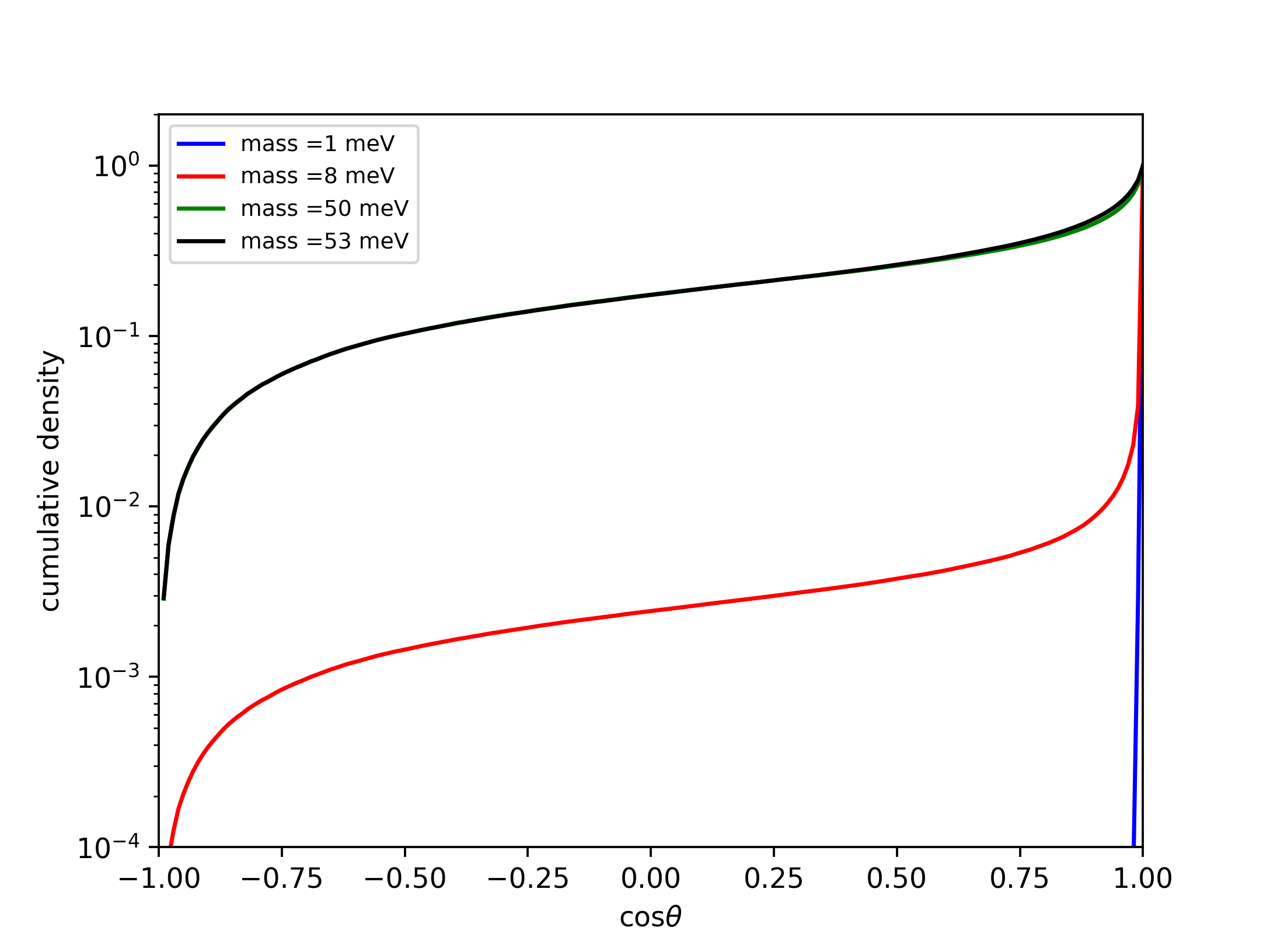}
\caption{Deflection angle distribution reconstruction calculated from neutrinos from the QUIJOTE simulation found inside haloes that satisfy $M>10^{15}\,M_\odot$.}
\label{fig:angle distribution reconstruction}
\end{figure}

In Fig.~\ref{fig:angle distribution reconstruction} we show the result of the reconstruction for the three neutrino mass eigenstates of our reference case (Eq.~\ref{eq:neutrino masses}). Also, we show the result obtained for the neutrinos of the used simulation that have a mass of 53 meV. It is clear that the less massive the neutrino is, the more relativistic is and the less deflection it suffers. Tab.~\ref{tab:angle cosine} shows the mean cosine of the deflection angle for neutrinos masses of reference and for both simulation suites: DEMNUni and QUIJOTE. The last was used in our previous work~\cite{Hernandez-Molinero:2022zoo}. The DEMNUni analysis presented here has 50 times more statistics than the previous analysis with QUIJOTE simulations; however, as it is shown in Tab.~\ref{tab:angle cosine}, results from QUIJOTE are compatible with the new results from DEMNUni, but it must be taken into account that we obtained the same results for an analysis constrained to haloes bigger than $10^{15}\,M_\odot$.

\begin{table}[H]
    \centering
\begin{tabular}{c c |*5{c}}
& mass [meV] & 1 & 8 & 33 & 50 & 53\\
\cline{2-7}
DEMNUni: & $\langle\cos\theta\rangle$ & 1.000 & 0.994 & 0.805\tablefootnote{Value obtained from fitting formula (Eq.~\ref{eq:linear fit}).} & 0.603 & 0.596\\
\cline{2-7}
QUIJOTE: & $\langle\cos\theta\rangle$ & 1.000 & 0.991 & 0.747 & 0.543 & \\
\end{tabular}
\caption{Mean cosine of the deflection angle for neutrinos inside big haloes ($M>10^{15}\,M_\odot$) for DEMNUni simulations and for the, lower resolution, QUIJOTE simulations.}
\label{tab:angle cosine}
\end{table}

\begin{figure}
\centering
\includegraphics[width=0.9\textwidth]{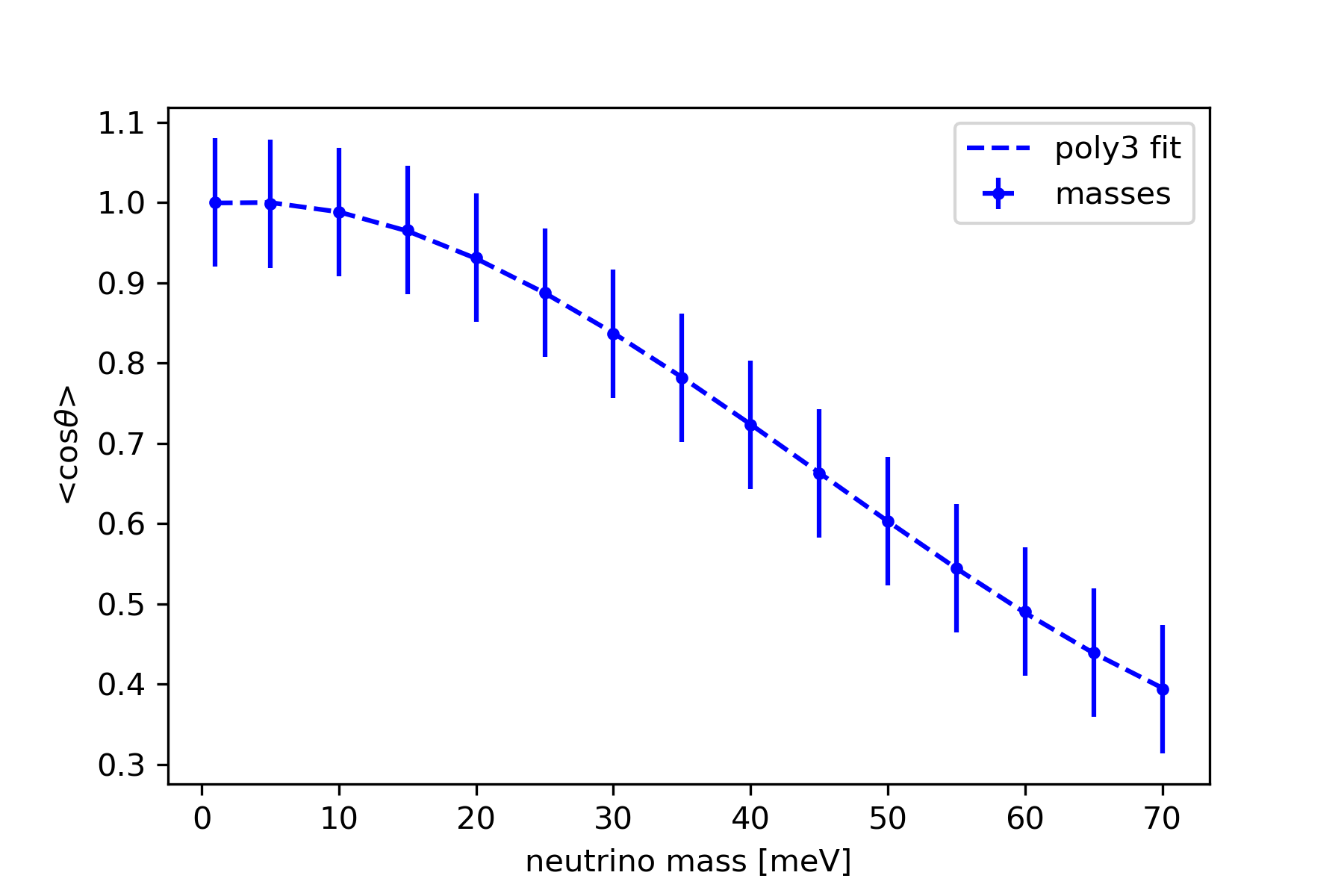}
\caption{Mean cosine of deflection angle vs. neutrino mass curve calculated for neutrinos from the DEMNUni simulation found inside haloes that satisfy $M>10^{15}\,M_\odot$. The error bar shows the typical $1\sigma$ uncertainty that will affect the angle distribution at each mass.}
 \label{fig:linear fit}
\end{figure}

Apart from neutrino masses of reference (Eq.~\ref{eq:neutrino masses}), this operation has been done for masses from 1 meV to 70 meV in steps of 5 meV. Once more, the mean cosine of the deflection angle is calculated for each mass eigenstate from the deflection angle distribution obtained from the simulation with neutrinos of total mass of 0.16 eV found in haloes within 1.5 Mpc that satisfy $M>10^{15}\,M_\odot$ (Fig.~\ref{fig:angle deflection distribution}). The result is presented in Fig.~\ref{fig:linear fit} as well as a polynomial fit. 

This result is a direct relation between the neutrino mass and the mean cosine of the deflection angle, deflection produced by dark matter haloes like the Local Supercluster.
\begin{equation}
    \langle\cos\theta\rangle_j = (2.30\,10^{-6} \text{meV}^{-3} \cdot m_{\nu_j}^3 -3.12\,10^{-4}\text{meV}^{-2} \cdot m_{\nu_j}^2 + 1.94\,10^{-3}\text{meV}^{-1} \cdot m_{\nu_j} + 0.998 ) \pm 0.08
    \label{eq:linear fit}
\end{equation}

The uncertainty in Eq.~\ref{eq:linear fit} comes from the scatter in the distribution of angles for the DEMNUni simulation with neutrino mass of $0.05$ eV and that is assumed for all other masses as an uncertainty estimation. This formula can be used whenever neutrino mass is not bigger than $80$ meV; in other words, fulfilling \ref{mass_condition1} and \ref{mass_condition2}, the lightest neutrino must have a mass smaller than $60$ meV.

\section{Capture rates}
\label{sec:capture}

The helicity changes of neutrinos crossing dark matter fields modify old calculations to compute neutrinos capture rates, especially for relativistic cases. Old calculations~\cite{PTolemy} established a factor of two for non-relativistic cases between Dirac and Majorana neutrinos capture rates. In our previous work~\cite{Hernandez-Molinero:2022zoo}, we established a factor of $1.91^{+0.09}_{-0.38}$ as a function of the unknown neutrino mass scale, we focused on the fact that the lightest neutrinos cannot be considered non-relativistic along with helicity changes by gravity. Now, we update this calculation taking into account that helicity densities depend on the neutrino mass eigenstate too and restricting our calculation to our solar environment.

Once more, the total capture cross section in a nucleus for a given neutrino mass and helicity multiplied by neutrino velocity (which is the relevant quantity for the capture rate) is given by
\begin{equation}
    \sigma_j(s_{\nu_j})v_{\nu_j} = |U_{ej}|^2\int_0^\infty{dq_\nu\Bar{\sigma}(q_\nu)(1-2s_{\nu_j}v_{\nu_j})}\,,
\end{equation}
where $\Bar{\sigma}(q_\nu)$ can be considered independent of the neutrino mass. $U_{ej}$ is the PMNS matrix element. And $s_\nu=-1/2$ for left-handed and $s_\nu=+1/2$ for right-handed neutrinos. Regarding capture rates what we had was
\begin{equation}
\Gamma_{C\nu B}= \left\{\left[\sum_{j=1}^{3}{\sigma_j(s_\nu = -1/2)v_{\nu_j}}\right]n(\nu_{hL}) + \left[\sum_{j=1}^{3}{\sigma_j(s_\nu = -1/2)v_{\nu_j}}\right]n(\nu_{hR})\right\}N_T  \,,
\end{equation}
where $n(\nu_{hL})$ and $n(\nu_{hR})$ were the same for all neutrinos masses eigenstates. However, now we are considering that left- and right-handed neutrinos densities are not the same for all neutrino mass eigenstates and, to calculate helicity densities for a specific neutrino mass eigenstate, the only thing needed is the mean cosine of the deflection angle which is given by Eq.~\ref{eq:linear fit}, i.e., neutrino densities can be calculated as follow
\begin{subequations}
    \begin{equation}
        n_j(\nu_{hL})=n_0\left[\frac{1}{2}+\frac{1}{2}\langle\cos\theta\rangle_j\right] \,,
    \end{equation}
    \begin{equation}
        n_j(\nu_{hR})=n_0\left[\frac{1}{2}-\frac{1}{2}\langle\cos\theta\rangle_j\right] \,.
    \end{equation}
    \label{eq:densities}
\end{subequations}

The fact that densities depend on masses lead to the following expression of capture rates
\begin{equation}
    \Gamma_{C\nu B}= \sum_{j=1}^{3}{\left[\sigma_j(s_\nu = -1/2)v_{\nu_j}n_j(\nu_{hL}) + \sigma_j(s_\nu = -1/2)v_{\nu_j}n_j(\nu_{hR})\right]}N_T  \,.
\end{equation}

Finally, Tab.~\ref{tab:rates} shows the result of the total cross section multiplied by the neutrino velocity for left- and right-handed neutrinos as well as the capture rate on 100g of tritium for the Dirac and Majorana cases for our masses of reference, as in Tab. 1 in~\cite{Hernandez-Molinero:2022zoo}. A new column is now added showing the densities of left- and right-handed neutrinos for each neutrinos masses eigenstates.

\begin{table}[H]
\makebox[1 \textwidth][c]{       
\resizebox{1 \textwidth}{!}{  
\begin{tabular}{c|c c c c a a}
\hline
\multirow{3}{*}{Neutrinos masses} & \multirow{2}{*}{$\sigma_j(\text{left})\cdot v_{\nu_j}$} & \multirow{2}{*}{$\sigma_j(\text{right})\cdot v_{\nu_j}$} & \multicolumn{2}{c}{Densities for Dirac cases} & \multicolumn{2}{a}{Capture rate} \\
& & & $n_j(\nu_{hL})$ & $n_j(\nu_{hR})$ & $\Gamma_{C\nu B}^D$ & $\Gamma_{C\nu B}^M$\\
& $[\times10^{-45}\,\text{cm}^2]$ & $[\times10^{-45}\,\text{cm}^2]$ & $[\text{cm}^{-3}]$ & $[\text{cm}^{-3}]$ & $[\text{yr}^{-1}]$ & $[\text{yr}^{-1}]$\\
\hline
$m_1 = 1\, \text{meV}$ & 3.6799 & 1.4458 & 55.9988 & 0.0012  & 3.89 & 5.46 \\
$m_2 = 8\, \text{meV}$ & 1.2261 & 1.0745 & 55.818 & 0.182  & 1.29 & 2.45 \\
$m_3 = 50\, \text{meV}$ & 0.0860 & 0.0842 & 44.881 & 11.119 & 0.09 & 0.18 \\
\rowcolor{black!15}       TOTAL & 4.9920 & 2.6045 &  &   & \textbf{5.28} & \textbf{8.09} \\
\end{tabular}
} 
} 
\caption{Summary of cross section, helicity state densities and capture rate results for our neutrino masses of reference.}
\label{tab:rates}
\end{table}

When neutrinos are very relativistic, the right-handed contribution to the total cross section is less than half compared to the left-handed contribution; so, even though the Majorana case has a right-handed contribution, it is much smaller than the left-handed one, equalling Dirac and Majorana results, leaving gravity aside. On the other hand, for non-relativistic cases, left- and right-handed contributions are almost equal, which means that gravity and neutrinos nature effects are more visible in these cases. However, only electron neutrinos contribute to the capture process and, if the hierarchy is normal, then the lightest mass eigenstate is the principal component of the electron neutrino, meaning that the highest contribution to the total capture rate is the one less affected by gravity clustering, so that gravity does not reduce so much the capture rate in the Dirac cases. The reduction of a factor of 2 between Majorana and Dirac cases that we are observing is because light neutrinos are relativistic and to a lesser extent due to gravity.

\section{Conclusions}
\label{sec:conclusions}

Gravity effects must be included in the $C_\nu B$ flux estimation of weakly interacting neutrinos in order to properly characterise their expected capture rate depending on their nature: Dirac or Majorana. Even though neutrinos are not virialised, they suffer a deflection when travelling inside our Local Supercluster. To overcome the problem of producing new cosmological simulations for different neutrino masses in order to calculate the deflection angle, we have developed a method that has allowed us to get a fitting formula providing the mean cosine of the deflection angle detected on Earth for any neutrino mass. With new helicity densities, we have updated our calculation of capture rates on tritium based detectors placed on the ground. 

Gravity produces changes in the helicity content as a function of neutrino masses: the lightest neutrinos do not suffer so much the deflection, whereas non-relativistic neutrinos are more affected by gravitational potentials; meaning that the relativistic behaviour of the neutrinos, along with gravity, modifies the prediction in case neutrinos are Dirac fermions. In general, gravity should decrease the capture rate but, in case neutrinos are very relativistic, gravity does not affect them so much, resulting in a milder suppression of the capture rate of Dirac neutrinos. And for non-relativistic neutrinos, left- and right-handed cross sections are equal, so, even though a deflection effect exists, it is not translated into the capture rate, having at the end a capture rate for Dirac half of Majorana neutrinos.

To summarise, the lightest neutrinos are so ultra-relativistic that they to do not suffer deflection from gravity but, at the same time, they are not relativistic enough to achieve the same capture rate for Dirac and Majorana cases. This means that the capture rate in Ptolemy-like experiments will be  sensitive to the neutrino nature, and gravity helps to increase that difference.

For future work, because the neutrino capture rate is dependent on the nature of gravity, if neutrinos are Dirac, it is tempting to speculate if their measured rate can be used to unveil the nature of gravity over scales comparable to the current horizon.

\begin{acknowledgments}
Funding for the work of RJ was partially provided by
project PGC2018-098866- B-I00 y FEDER “Una manera
de hacer Europa”, and the “Center of Excellence Maria de Maeztu 2020-2023” award to the
ICCUB (CEX2019- 000918-M) funded by MCIN/AEI/10.13039/501100011033. 
The DEMNUni simulations were carried out in the framework of ``The Dark Energy and Massive-Neutrino Universe" project, using the Tier-0 IBM BG/Q Fermi machine and the Tier-0 Intel OmniPath Cluster Marconi-A1 of the Centro Interuniversitario del Nord-Est per il Calcolo Elettronico (CINECA). We acknowledge a generous CPU and storage allocation by the Italian Super-Computing Resource Allocation (ISCRA) as well as from the coordination of the ``Accordo Quadro MoU per lo svolgimento di attività congiunta di ricerca Nuove frontiere in Astrofisica: HPC e Data Exploration di nuova generazione'', together with storage from INFN-CNAF and INAF-IA2.
\end{acknowledgments}

\bibliographystyle{JHEP}

\providecommand{\href}[2]{#2}\begingroup\raggedright\endgroup

\end{document}